\documentclass[%
 reprint,
superscriptaddress,
 amsmath,amssymb,
 aps,
 prl,
]{revtex4-2}

\usepackage{aas_macros}

\usepackage{xcolor}
\definecolor{revblue}{RGB}{0,70,180}
\def\bt{\textcolor{black}}
\def\BT{\textcolor{black}}

\usepackage{placeins}
\usepackage{graphicx}% Include figure files
\usepackage{dcolumn}% Align table columns on decimal point
\usepackage{bm}% bold math
\usepackage[colorlinks=true,allcolors=blue]{hyperref}
\usepackage{xcolor}

\begin{document}

\title{Exponential Growth of Turbulent Cross Helicity in Shear-Flow-Driven Turbulence}

\author{B.~Tripathi}
\email{Email: bt2693@columbia.edu}
\affiliation{Columbia University, New York 10027, USA.}
\affiliation{University of Wisconsin--Madison, Madison, Wisconsin 53706, USA.}
\author{A.~E.~Fraser}
\affiliation{University of Colorado, Boulder, Colorado 80309, USA.}
\author{P.~W.~Terry}
\author{E.~G.~Zweibel}
\affiliation{University of Wisconsin--Madison, Madison, Wisconsin 53706, USA.}
\author{M.~J.~Pueschel}
\affiliation{Dutch Institute for Fundamental Energy Research, 5612 AJ Eindhoven, The Netherlands.}
\affiliation{Eindhoven University of Technology, 5600 MB Eindhoven, The Netherlands.}
\affiliation{Department of Physics \& Astronomy, Ruhr-Universit\"at Bochum, 44780 Bochum, Germany.}
\author{A.~M.~Beloborodov}
\affiliation{Columbia University, New York 10027, USA.}
\date{\today}

\begin{abstract}

\bt{Cross helicity, the velocity--magnetic-field correlation, is an ideal invariant of incompressible magnetohydrodynamics and constrains energy cascades and magnetic-field generation. Although cross helicity typically decays via visco-resistive effects, we analytically demonstrate that it grows exponentially in sheared inhomogeneous turbulence. This {\em cross-helical shear instability} is confirmed using three-dimensional numerical simulations. Growth requires a nonzero gradient of the Reynolds--Maxwell stress to produce turbulent cross helicity; this production overcomes effective losses when shear exceeds a threshold set by the ratio of turbulent magnetic to momentum diffusivities.}

\end{abstract}

\maketitle

\bt{Cross helicity, the correlation between flow and magnetic field, plays a central role in magnetohydrodynamic (MHD) turbulence, regulating energy transfer, turbulent transport, and magnetic-field amplification \cite{chandrasekhar1961,dobrowolny1980,politano1998,perez2009,yokoi1999,yokoi2013,yokoi2023,beattie2026}.
The generation of cross helicity is, however, nontrivial.
In a homogeneous system, cross helicity may be locally generated, but its volume average is ideally conserved and depleted by visco-resistive effects \cite{dobrowolny1980}. 
Existing mean-field theories describe how an already appreciable cross helicity modifies MHD  dynamics \cite{yoshizawa1990,yokoi1999,yokoi2011,yokoi2013,yokoi2023,blackman1997,brandenburg1998,blackman2000,blackman2002,higashimori2013,yokoi2011b,widmer2019,yokoi2013PhPl}. 
For instance, a key role of cross helicity was recently observed in numerical simulations, which showed that large-scale magnetic fields can be produced through the mean-vorticity effect, once a significant cross helicity is present \cite{tripathi2026nature}. 
However, this body of work does not explain how such an appreciable cross helicity is produced in the first place. The central question addressed here is, therefore, under what conditions cross helicity overcomes damping and grows {\em exponentially}.}

\bt{Flow shear provides a natural setting for this question because it anisotropizes velocity and magnetic fluctuations and can correlate the two \cite{blackman1997,blackman2000,rogachevskii2003,rogachevskii2004}. Previous analytic treatments considered shear $S$  as a weak perturbation to prescribed turbulence, $S=\tau|\nabla\mathbf U|\ll1$, where $\tau$ is the turbulence correlation time and $\mathbf U$ denotes a large-scale flow \cite{radler2003,rogachevskii2003,rogachevskii2004,kleeorin2008,rogachevskii2008}. Rogachevskii and Kleeorin (RK) showed  that weak shear and turbulence inhomogeneity produce velocity--magnetic-field correlations that depend linearly on the velocity gradient \cite{rogachevskii2003}. 
For $S\gtrsim1$, however, shear can itself generate or modify  the turbulence. In this Letter, we demonstrate that sufficiently strong shear enables turbulent cross helicity to grow exponentially via an instability which is absent when $S\ll 1$.}

\bt{Cross helicity in MHD fluctuations, also called turbulent cross helicity, shares a useful analogy with turbulent kinetic energy in hydrodynamics: both obey budget equations containing production, transport, and dissipation \cite{pope2000,yokoi2011,yokoi2013,yokoi2023}. In hydrodynamic shear turbulence, the evolution of turbulent kinetic energy involves the Reynolds stress acting against the mean-flow gradient \cite{pope2000}. 
Similarly, in MHD, alongside the effect just mentioned, the evolution of turbulent cross helicity involves the combination of the Reynolds and Maxwell stresses acting against the gradient of the mean magnetic field \cite{yokoi2011,yokoi2013,yokoi2023}; in addition to this ``stress-production'' term $\mathcal P_{\rm stress}$, MHD equations also give an ``EMF transfer'' term $\mathcal P_{\rm EMF}$ describing the interaction of the turbulent electromotive force (EMF) with the shear \cite{yokoi2011,yokoi2013,yokoi2023}. Unlike the hydrodynamic term, these MHD terms are not sign-definite, even for a given shear of fixed sign. Thus, it is unknown {\em a priori} whether they produce or deplete turbulent cross helicity. Identifying the net source, the losses, and the region where production exceeds losses is, therefore, essential for determining the conditions under which turbulent cross helicity can grow or decay.}

\bt{The basic picture of the cross-helical shear instability (CHSI), confirmed later, is as follows: The volume-averaged turbulent cross helicity $W$ evolves as
\begin{equation}
\partial_t W=\mathcal P_{\rm stress} + \mathcal P_{\rm EMF} + \mathcal D,
\label{eq:Upsilon_intro}
\end{equation}
where $\mathcal D$ relates to visco-resistive dissipation. 
An initial infinitesimal $W$ generates a small mean magnetic field via the electromotive force, rendering $\mathcal P_{\rm EMF}$ negative relative to $W$; see Eq.~(S8) in the Supplement and Figs.~\ref{fig:f4_new} and \ref{fig:f3_new_round2}.
This mean field enters the stress-production $\mathcal P_{\rm stress}$, thereby providing feedback on $W$. We find $\mathcal P_{\rm stress}$ to be positive relative to $W$.  Using three mean-field derivations---given below and in the End Matter---we show that growth requires two ingredients: a nonzero Reynolds--Maxwell-stress gradient to supply the source $\mathcal P_{\rm stress}$, and a finite shear for this source to overcome losses via turbulent EMF $\mathcal P_{\rm EMF}$ and visco-resistive dissipation $D$. This stress-production effect is subdominant in weak shear. The critical shear is set by $(\beta/\mu_T)^{1/2}$, where $\beta$ and $\mu_T$ are turbulent magnetic and momentum diffusivities.}

\textit{Mean-field formulation.---}
Let a suitable ensemble average define large-scale flow
\(\mathbf U=\langle\mathbf u_{\rm tot}\rangle\) and large-scale magnetic field
\(\mathbf B=\langle\mathbf b_{\rm tot}\rangle\), and fluctuations
\(\mathbf u=\mathbf u_{\rm tot}-\mathbf U\) and
\(\mathbf b=\mathbf b_{\rm tot}-\mathbf B\), where $\mathbf u_{\rm tot}$ and $\mathbf b_{\rm tot}$ are the total flow and magnetic field. The mean induction equation is
\begin{equation}\label{eq:mean_general}
\partial_t\mathbf B
=
\nabla\times(\mathbf U\times\mathbf B)+\nabla\times\boldsymbol{\mathcal E}
+\eta\nabla^2\mathbf B,\qquad \boldsymbol{\mathcal E} = \langle\mathbf u\times\mathbf b\rangle,
\end{equation}
where $\boldsymbol{\mathcal E}$ is the turbulent electromotive force (EMF).

\textit{Turbulent-cross-helicity evolution.---}  This work focuses on the conditions for the growth of turbulent cross helicity $W=\langle\mathbf u\cdot\mathbf b\rangle_V$, where $\langle \cdot\rangle_V$ denotes volume average.  To write the evolution equation for $W$, we subtract the evolution equation for the mean-field cross helicity $\langle \mathbf U \cdot \mathbf B\rangle_V$ from that of the total cross helicity $\langle \mathbf u_{\rm tot} \cdot \mathbf  b_{\rm tot}\rangle_V$, 
\begin{equation}\label{eq:turbcrosshel}
\partial_t  W \mathrm{=} 
\hspace{0.2cm}\underbrace{\langle \mathbf{B}\cdot\mathbf{T}\rangle_V}_{\textrm{\hspace{-1.0cm}}\substack{{\mathcal P}_{\rm stress}:\\ \text{\BT{stress-production}}\\ \text{\BT{}}}\hspace{-1cm}} 
-(\nu+\eta)\langle\boldsymbol{\omega}\cdot\mathbf{j}\rangle_V 
\underbrace{- \langle \mathbf{U}\cdot\nabla\times\boldsymbol{\mathcal{E}}\rangle_V}_{\textrm{\hspace{-1.0cm}}\substack{{\mathcal P}_{\rm EMF}:\\ \text{\BT{EMF transfer}}\\ \text{\BT{}}}\hspace{-1cm}} ,
\end{equation}
where $T_i \mathrm{=} \partial_j R_{ij}$ and $R_{ij}\mathrm{=}\langle u_i u_j-b_i b_j\rangle$ is the Reynolds--Maxwell stress \cite{yoshizawa1999,yokoi2013}. The dissipative term is written in terms of $\boldsymbol{\omega}\mathrm{=}\nabla\times\mathbf{u}$ and $\mathbf{j}\mathrm{=}\nabla\times\mathbf{b}$ in a periodic or closed system;  $\nu$ is the  kinematic viscosity and $\eta$ is the magnetic diffusivity. Equation~\eqref{eq:turbcrosshel}, derived in Supplementary Information (SI), is the exact volume-averaged turbulent cross helicity budget for incompressible MHD. No direct external forcing is applied to fluctuations.

The first term on the right-hand side of Eq.~\eqref{eq:turbcrosshel} acts as a source for $W$ when this term and $W$ have the same sign \cite{yokoi2013, yokoi2023}. \bt{After integration by parts,
\(\langle\mathbf B\cdot\mathbf T\rangle_V
\mathrm{=}-\langle R_{ij}\partial_jB_i\rangle_V\).}
Similarly, the term $- \langle \mathbf{U}\cdot\nabla\times\boldsymbol{\mathcal{E}}\rangle_V\mathrm{=} -\langle \boldsymbol{\mathcal{E}}\cdot(\nabla\times\mathbf{U})\rangle_V$, when negative relative to $W$, 
represents transfer of cross helicity from fluctuations to the mean fields.

\textit{Dispersion relation.---} 
To isolate the role of cross helicity, we separate the direct mean-flow induction operator
\(\nabla\times(\mathbf U\times\mathbf B)\) of Eq.~\eqref{eq:mean_general}.  When the ensemble average is suitably chosen such that it averages turbulent correlations over two statistically homogeneous directions, thereby defining mean as a quantity that varies along the third dimension, \(\nabla\times(\mathbf U\times\mathbf B)\) may be set to zero. This is the case for the numerical simulations we consider below. While this consideration makes the analysis simpler, much of what follows can be applied to more general systems, with appropriate modifications.

\bt{To solve Eq.~\eqref{eq:turbcrosshel}, an expression for $\boldsymbol{\mathcal{E}}$ is required. Full tensorial structure of $\boldsymbol{\mathcal{E}}$ is given in the End Matter, where anisotropy and inhomogeneity of transport coefficients are retained and their connection to the weak-shear RK tensor basis is shown.} Here, for analytic consideration, we specialize in the case where turbulent cross helicity acts as the main source in the EMF \cite{yokoi2013, yokoi2023,yokoi1999,yokoibalarac2011,yokoi2011b,widmer2019,yokoi2013PhPl}: $\boldsymbol{\mathcal E}
\simeq \Upsilon\nabla\times\mathbf U-\beta\nabla\times\mathbf B,$
where $\Upsilon=2\tau W /3$ and \(\beta\) is the turbulent magnetic diffusivity. With this expression for $\boldsymbol{\mathcal E}$ and the direct induction operator separated, Eq.~\eqref{eq:mean_general} becomes $[\partial_t -(\beta+\eta) \nabla^2 ] \mathbf B = \nabla\times[(2\tau W/3)\nabla\times\bf U]$, whose Fourier transform \(\partial_t\to\lambda\) gives
\begin{equation}\label{eq:BUrelation_3D}
\widehat{\mathbf B}(\mathbf k)
=
\frac{k^2\widehat{\mathbf U}(\mathbf k)}
{\lambda+\beta_T k^2} \frac{2\tau W}{3},
\qquad
\beta_T=\beta+\eta,
\qquad
k=|\mathbf k|,
\end{equation}
where $\widehat{\mathbf B}(\mathbf k)$ and $\widehat{\mathbf U}(\mathbf k)$ are Fourier harmonics; \bt{the dominant spatial dependence is carried by the shear profile and by the mean magnetic field, as they tend to vary on a similar scale when the EMF is of the form shown above.} In the kinematic phase,  magnitudes of $W$ and $\widehat{\mathbf{B}}$ evolve as $e^{\lambda t}$, where $\lambda$ is to be determined. A mean flow $\mathbf U$ is prescribed.

Substituting Eq.~\eqref{eq:BUrelation_3D} and the EMF expression in Eq.~\eqref{eq:turbcrosshel},
\begin{equation}\label{eq:dispersion}
\begin{aligned}
\lambda+ c_{\rm damp}(\lambda)= \frac{2\tau}{3}  \sum_{\mathbf{k}}\,\,\, \left[\frac{k^2 \widehat{\mathbf{U}}^\ast(\mathbf{k})\cdot \widehat{\mathbf{T}}(\mathbf{k})}{\lambda +\beta_\mathrm{T} k^2}\right],
\end{aligned}
\end{equation} 
where the term on the right-hand side corresponds to the stress-production term of Eq.~\eqref{eq:turbcrosshel}; see Sec.~S3 in SI for a step-by-step derivation. The symbol $\widehat{\mathbf{U}}^\ast$ denotes the complex conjugate of $\widehat{\mathbf{U}}$. The scalar \(c_{\rm damp}(\lambda)\) denotes the effective losses due to visco-resistive effects and EMF transfer, which are the last two terms of Eq.~\eqref{eq:turbcrosshel}, 
\begin{equation}\label{eq:cdamp}
c_{\rm damp}(\lambda) = \tau_{\rm diss}^{-1}+\frac{2\tau}{3}  \sum_{\mathbf{k}}\,\,\, \left[\frac{k^2 |\widehat{\mathbf{U}}(\mathbf{k})|^2 (\lambda +\eta k^2)}{\lambda +\beta_\mathrm{T} k^2}\right].
\end{equation} 
In Eq.~\eqref{eq:cdamp},  the visco-resistive term is written as
\((\nu+\eta)\langle\boldsymbol\omega\cdot\mathbf j\rangle_V
=\langle\mathbf u\cdot\mathbf b\rangle_V/\tau_{\rm diss}\),
which defines the effective cross-helicity damping time
\(\tau_{\rm diss}\). For real $\lambda>0$, \(c_{\rm damp}\) is non-negative. Thus, if $\widehat{\mathbf T}=0$, Eq.~\eqref{eq:dispersion} has no growing root.
Hence, a growing solution of Eq.~\eqref{eq:dispersion} requires the stress-production term on the right-hand side of Eq.~\eqref{eq:dispersion} to be non-zero and to exceed the effective loss rate $c_{\rm damp}$. 

\textit{Critical shear threshold.---}
The stress gradient $\widehat{\mathbf T}$ in Eq.~\eqref{eq:dispersion} is nonzero when turbulence is inhomogeneous, which can arise due to external inhomogeneity or a non-uniform shear. For analytic clarity, take the representative harmonic $\mathbf U=U_0\sin(Kz)\hat{\mathbf x}$; adding harmonics to represent the simulated tanh shear gives the same qualitative growing branch. Equation~\eqref{eq:dispersion} then reduces to
\begin{equation}\label{eq:lambdaeqn}
\lambda
=
-D
\pm
\sqrt{
D^2+(\tau/3)(U_0K^2)^2
\left[
\mu_T
-\eta_\ast
\right] } ,
\end{equation}
where turbulent-stress gradient produces the term with $
\mu_T= 2\mathrm{Im}{(-\widehat T_x(K))}/(U_0K^2)$; here, \(T_x = \partial_z \langle u_xu_z\rangle=-\partial_z (\mu_T\partial_zU_x)\).  Damping of turbulent cross helicity arises from
\(\eta_\ast=\eta+3\beta_T/(\tau U_0^2K^2\tau_{\rm diss})\) and \(D\mathrm{=}[\beta_T K^2+\tau U_0^2K^2/3+\tau_{\rm diss}^{-1}]/2\). In the absence of flow shear $K\mathrm{=}0,$ Eq.~\eqref{eq:lambdaeqn} lacks a growing solution. In the presence of flow shear, when $
\mu_T >\eta_\ast$, Eq.~\eqref{eq:lambdaeqn} admits a growing solution. This positive branch is the CHSI.

Estimating \(\tau_{\rm diss}=\tau/c_d\) where $c_d$ is a constant \cite{yoshizawa1990,yokoi2023}, the criterion \(\mu_T>\eta_\ast\) reduces to \bt{a criterion for the shear rate $S=\tau|\nabla\mathbf U|= \tau U_0 K$ in the ideal limit ($\eta=0$),
\begin{equation}\label{eq:Sconditon}
S \gtrsim \sqrt{\frac{3 c_d \beta}{\mu_T}}.
\end{equation}
The factor $(3 c_d \beta/\mu_T)^{1/2}$ is generally expected to be of the order unity. Thus, in the weak-shear regime \(S\ll1\), the stress-production effect, though present, is overpowered by turbulent and dissipative losses.}

\bt{The above derivation of the CHSI criterion \eqref{eq:Sconditon} is our first mean-field derivation. This criterion is reproduced by two additional mean-field derivations, given below before the Discussion and in the End Matter. Those two derivations  retain the spatial inhomogeneity of $W$, $\beta$, and related coefficients, as well as the full anisotropy of shear-flow turbulence; see Eqs.~\eqref{eq:QLestimate} and \eqref{eq:etaeff_threshold_EM}.}

\textit{Numerical simulations.---} We numerically simulate three-dimensional (3D) incompressible MHD turbulence \cite{chandrasekhar1961} where the $(x,y)$-averaged mean magnetic field $\mathbf B$ is initially zero.  We impose only one large-scale ingredient: a maintained mean shear flow $U_x(z)\mathrm{=}U_0 \left\{\tanh[(z-z_1)/a] - \tanh[(z-z_2)/a] -1 \right\}$, where $U_0$ is the flow amplitude 
and $a$ is the half-width of the shear layers \cite{lecoanet2016} located at $z_1\mathrm{=}L_z/4$ and $z_2\mathrm{=}3 L_z/4$ (flow-reversal locations) with $L_z$ the extent of the domain in $z$. Length and time scales are measured in units of $a$ and $a/U_0$, respectively.
Since the analytic theory considers a prescribed large-scale shear flow, we test the theory by maintaining the large-scale shear flow in simulations. Findings are qualitatively similar in maintained shear flows whether the shear flow is frozen (a standard procedure in dynamo studies \cite{yousef2008,squire2015}) or the instantaneous shear flow $U_x(z,t)$ is dynamically driven toward its initial unstable equilibrium \cite{smith2021, marston2008, cope2020, pueschel2011, pueschel2014}.  \bt{In these runs, the Kelvin--Helmholtz instability (KHI) of the flow $U_x(z)$ drives turbulence, which naturally operates with $S=\tau |\nabla\mathbf U|\gtrsim 1$; see Fig.~\ref{fig:shear_rate}.}

Since the unstable flow $U_x(z)$ drives turbulence that is \bt{statistically homogeneous in $x$ and $y$ and is statistically inhomogeneous in $z$ (see Fig.~\ref{fig:homogeneity_test})}, we define \textit{mean} as the plane average over the statistically homogeneous $x$ and $y$ directions; this average is denoted by $\langle \cdot\rangle$ hereafter. In such a case, the direct induction $\nabla\times(\mathbf U \times \mathbf B) = 0$, as $U_z=B_z=0$ is used in the simulations for the $(x,y)$-averaged mode ($\nabla\cdot\mathbf U=\nabla\cdot\mathbf B=0$).

On top of the mean shear flow $U_x(z)$, we add phase-randomized, broadband, divergence-free, fully 3D fluctuations in velocity and magnetic fields, whose amplitudes compared to that of the mean flow are $\mathrm{\sim}10^{-9}$. No direct external force is applied to fluctuations, and they are evolved self-consistently using MHD equations.

 \begin{figure}
    \centering
    \includegraphics[width=1\columnwidth]{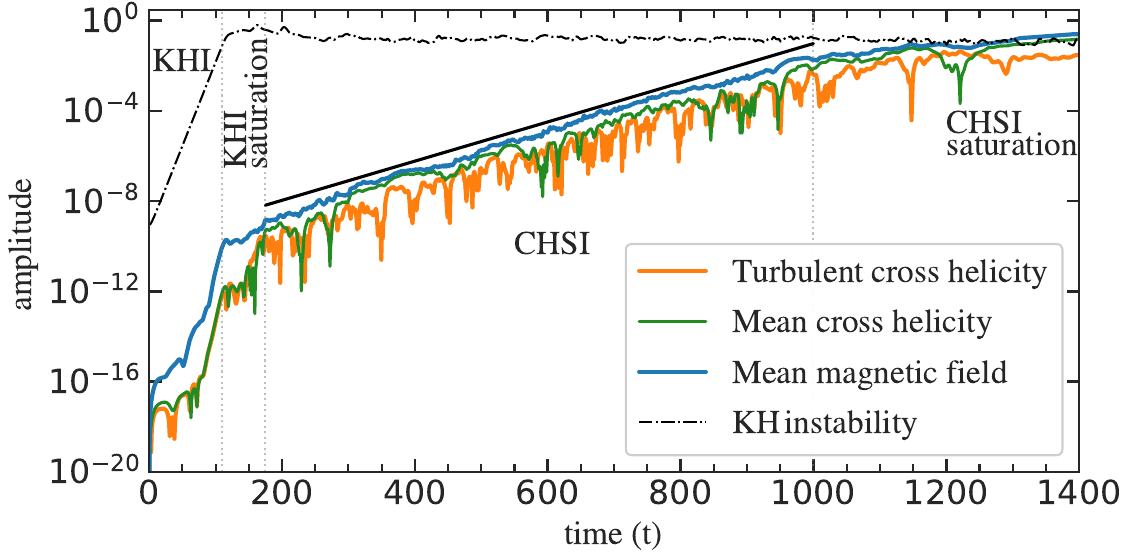}
    \caption{Evolution of turbulent cross helicity and mean magnetic field. The vertical axis shows the absolute value of each curve. Phase 1: KHI grows; the fastest growing mode is shown. Phase 2: KHI saturates and becomes fully 3D hydrodynamically (negligible Lorentz force at this stage). This alters the cross-helicity evolution. Phase 3: CHSI dominates; $\langle\mathbf u\cdot\mathbf b\rangle_V$ and the mean magnetic field grow with the same exponential rate. Phase 4: saturation occurs when cross-helicity production balances losses. Parameters used are $\nu/(U_0a)=0.0125$ and $\eta/(U_0a)=0.02$. The spatial resolution is $512\times 512\times 1024$.} 
    \label{fig:f2_new}
\end{figure}

The evolution of turbulent cross helicity and mean magnetic field is shown in Fig.~\ref{fig:f2_new}. In phase 1, the unstable mean shear flow drives the KHI. Thus, both  two-dimensional and three-dimensional fluctuations with $k_x\neq 0$ grow exponentially by a linear process.  Upon nonlinear saturation of the hydrodynamic KHI, linearly stable wavenumbers $k_x=0$ and $k_y\neq 0$ are nonlinearly excited to large amplitudes. These fluctuations induce turbulent EMF, producing the plane-averaged mean field \cite{tripathi2026nature} ($k_x\mathrm{=}k_y\mathrm{=}0$). \bt{This, however, is not sufficient to explain the growth of turbulent cross helicity, which is the focus of this work.}

\begin{figure}
    \centering
    \includegraphics[width=0.9\columnwidth]{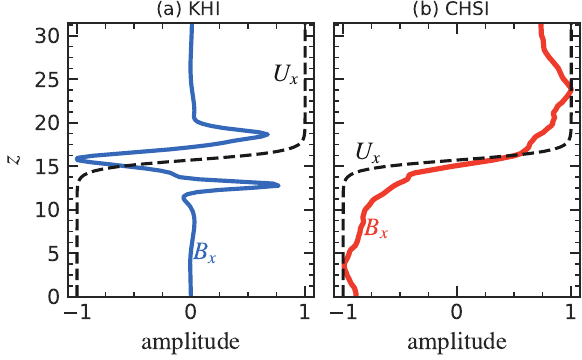}
    \caption{\bt{Mean-field profiles in the KHI phase (a) and the slowly growing CHSI phase (b). In the KHI phase, $B_x(z)$ is localized near the shear layer and can depend on initial perturbations. In the slow-growth CHSI phase, $B_x(z)$ follows the imposed $U_x(z)$, as predicted by Eq.~\eqref{eq:BUrelation_3D} when $\lambda/(\beta_Tk^2)\ll1$.  The profiles are time-averaged and separately normalized for comparison. Since  $B_x$ in the CHSI phase can change sign, that sign is taken into account while time-averaging.  The simulation uses $\eta/(U_0a)=0.03$ and zero initial mean magnetic field.}}
    \label{fig:mean_flow_profile_eta_0p03}
\end{figure}

The kinematic mean-field eigenmode is localized near the shear layers in phase 1, transitional in phase 2, and spatially extended in phase 3. For $\lambda/(\beta_T k^2)\ll1$, Eq.~\eqref{eq:BUrelation_3D} predicts nearly matching large-scale mean-field and mean-flow profiles. This limit occurs at CHSI saturation ($\lambda\approx0$) and can also be accessed in slowly growing kinematic phase. We obtain the latter regime by increasing the resistivity $\eta$ alone until $\lambda$ is substantially reduced; all other parameters are fixed. These controlled runs retain the full turbulent flow while reducing small-scale-dynamo contamination of the magnetic eigenfunction. Figure~\ref{fig:mean_flow_profile_eta_0p03} compares $B_x$ and $U_x$.

The CHSI growth rate is largely unchanged when the domain size $L_y$ is changed from $10\pi$ to $20\pi, 40\pi, 80\pi, 160\pi,$ and $320\pi$. See Fig.~\ref{fig:f5_app}.

Figure~\ref{fig:f4_new} shows that stress production, primarily $\langle B_xT_x\rangle$, drives $W$, while EMF transfer to the mean fields and visco-resistive effects damp it on average. Instantaneous values  rapidly fluctuate, but applying a low-pass filter shows a clear signal, consistent with the unfiltered data in Fig.~\ref{fig:f4_new}.

\bt{Figure~\ref{fig:f3_new_round2} shows the spatial maps of production, loss, and transport mechanisms using the spatial cross-helicity budget, which is $(x,y)$-averaged \cite{yokoi2013}: $\partial_t \langle \mathbf u\cdot \mathbf b\rangle(z) = \mathcal P_{\rm stress}(z) + \mathcal P_{\rm EMF}(z) + \mathcal T(z) + \mathcal D(z)$, where stress production is $\mathcal P_{\rm stress}(z)=-R_{iz}\partial_z B_i$, EMF transfer is $\mathcal P_{\rm EMF}(z) = -\mathcal{E}_i (\nabla\times \mathbf{U})_i$, dissipation is $\mathcal D(z) = \nu \langle b_i \partial_{jj} u_i \rangle + \eta \langle u_i \partial_{jj} b_i \rangle$, and transport is $\mathcal T(z) = \partial_z \mathcal T'$, with $\mathcal T'=\langle b_z (u^2/2 +b^2/2 -p_\ast) -u_z (u_j b_j) \rangle$, and $p_\ast$ combines the fluid and magnetic pressure; repeated indices are summed over.}

\bt{\textit{A quasilinear mean-field derivation.---}
We consider an analogy with hydrodynamics where vertical turbulent motions $u_z$ advect the mean horizontal momentum $U_x(z)$, producing horizontal fluctuations $u_x\simeq-\tau u_z\partial_zU_x$, as shown in Fig.~\ref{fig:sch}. Similarly, vertical turbulent motions $u_z$ advect the turbulence-generated mean horizontal magnetic field $B_x(z)$, producing $b_x\simeq-\tau u_z\partial_zB_x$. This magnetic response gives one contribution to the local velocity--magnetic correlation, $\langle u_xb_x\rangle\simeq-\tau\langle u_xu_z\rangle\partial_zB_x$, identifying the Reynolds part of the stress-production source.  The full Reynolds--Maxwell stress and the EMF-transfer terms are shown next.}

\bt{The full contribution to $\langle u_xb_x\rangle$ is estimated by iteratively replacing quasilinear expressions for $u_x$ and $b_x$: $\langle u_xb_x\rangle\simeq\langle u_x b_x^{(1)} + u_x^{(1)} b_x \rangle$, where $b_x^{(1)}\simeq\tau[-u_z\partial_zB_x+b_z\partial_zU_x+B_j\partial_ju_x]$ and $u_x^{(1)}\simeq\tau[-u_z\partial_zU_x+b_z\partial_zB_x+B_j\partial_jb_x]$. Hence, $
\langle u_xb_x\rangle\simeq
-\tau R_{xz}\partial_zB_x
-\tau\mathcal E_y\partial_zU_x,$}
\bt{where $R_{xz}=\langle u_xu_z-b_xb_z\rangle$. The term with $B_j$ drops out as $\langle \cdot\rangle$ represents $(x,y)$-average. This estimate recovers the stress-production and EMF-transfer terms of Eq.~\eqref{eq:turbcrosshel}.}

 \begin{figure}
    \centering
    \includegraphics[width=1\columnwidth]{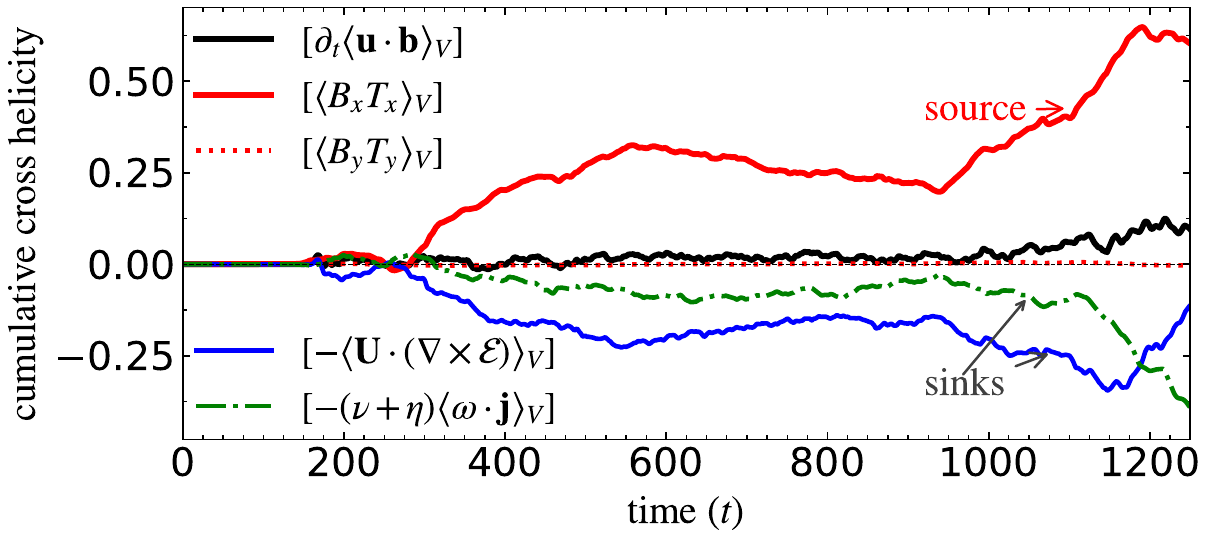}
    \caption{Temporal analysis of sources and sinks in the turbulent cross-helicity budget \eqref{eq:turbcrosshel}. Since magnetic fields grow exponentially in the kinematic phase, each term of Eq.~\eqref{eq:turbcrosshel} is divided by the magnetic rms amplitude \(b_{\rm rms}(t)\) and time-integrated, $[Q] = \int_0^t dt'\, Q(t')/b_{\rm rms}(t')$, to assess the cumulative generation or loss of turbulent cross helicity via different mechanisms. The CHSI phase begins at $t\mathrm{\approx}200$. The turbulent cross helicity $\langle \mathbf u\cdot \mathbf b\rangle_V$ is generated by the stress-production effect, in particular $\langle B_x T_x\rangle_V$. Since the source is generally larger than the sinks,  the value of the black curve, though small, is almost always positive. This positive value provides the growth of $\langle\bf u\cdot \bf b\rangle_V$ observed in Fig.~\ref{fig:f2_new}.}
    \label{fig:f4_new}
\end{figure}

\bt{Similar to the preceding paragraph, the EMF $\mathcal E_y=\langle u_zb_x-u_xb_z\rangle$ can be written as $\mathcal E_y \simeq-\beta_z\partial_zB_x+2S\langle u_zb_z\rangle$, where $\beta_z=\tau\langle u_z^2+b_z^2\rangle$ and $S=\tau\partial_zU_x$. Considering $\langle u_zb_z\rangle\simeq\langle u_xb_x\rangle$ and substituting the previous relation for $\langle u_xb_x\rangle$ gives $\mathcal E_y\simeq-\eta_{\rm eff}(z)\partial_zB_x$, with
\begin{equation}\label{eq:etaeffPRL}
\eta_{\rm eff}(z)=
\frac{C_{\rm QL} \beta_z+\tau S R_{xz}}{C_{\rm QL}+S^2}
\simeq
\frac{C_{\rm QL} \beta_z-\mu_TS^2}{C_{\rm QL}+S^2},
\end{equation}}
\bt{where $R_{xz}\simeq-\mu_T\partial_zU_x$ (see Fig.~S3) and $C_{\rm QL}$ is a constant that depends on the closure model; for the quasilinear model considered, $C_{\rm QL}=1/2$. In the absence of shear $S=0$, $\eta_{\rm eff}$ reduces to the ordinary turbulent diffusivity $\beta_z$. A similar expression for $\eta_{\rm eff}(z)$ is derived in the End Matter using tensor algebra and modeling nonlinear effects.}

\bt{The plane-averaged induction equation is $(\partial_t-\eta\partial_z^2)B_x=-\partial_z\mathcal E_y = \partial_z [\eta_{\rm eff}(z)\partial_z B_x]$. When $\eta_{\rm eff}(z)$ is negative in a strongly sheared region of space, $B_x$ may grow. Hence, using $\eta_{\rm eff}<0$, one may write a local condition for the growth of $B_x$ --- and the accompanying growth of $\langle \mathbf u \cdot \mathbf b \rangle_V$ --- which in the ideal limit becomes
\begin{equation}
S^{\rm QL}\gtrsim\sqrt{\frac{C_{\rm QL} \beta_z}{\mu_T}}.
\label{eq:QLestimate}
\end{equation}}
\bt{This threshold is expected to be not far from unity.}

\bt{Importantly, the destabilizing stress-production effect enters as $\mathcal O(S^2)$ in Eq.~\eqref{eq:etaeffPRL}: this effect is subdominant for $S\ll1$, but can overcome losses for $S\gtrsim1$. The ratio of turbulent magnetic  to momentum diffusivities seen in Eq.~\eqref{eq:QLestimate} is reproduced by all three mean-field derivations; the pre-factor like $C_{\rm QL}$ depends on the closure model. See Eqs.~\eqref{eq:Sconditon} and \eqref{eq:etaeff_threshold_EM}.}

\begin{figure}
    \centering
    \includegraphics[width=0.9\columnwidth]{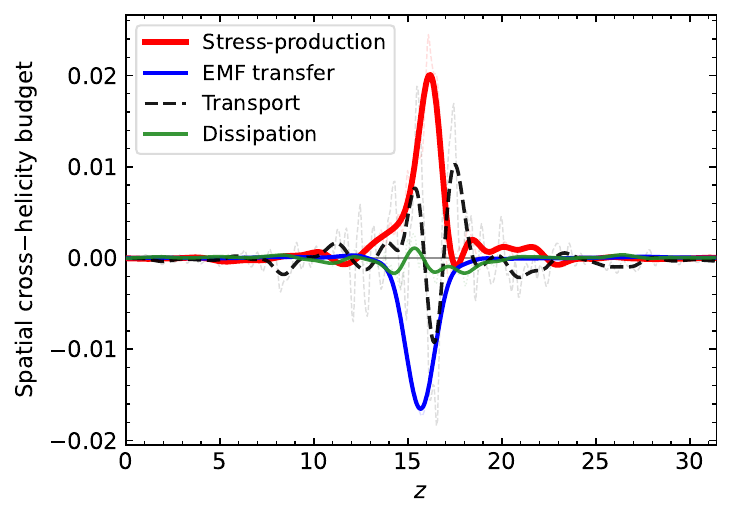}
    \caption{\bt{Spatial turbulent-cross-helicity budget during the kinematic CHSI phase. Turbulent cross helicity is produced in the shear layer by the Reynolds--Maxwell stress, while EMF transfer and visco-resistive dissipation act as losses; transport redistributes cross helicity without changing its volume average. Curves are time-averaged after normalizing them by $b_{\rm rms}$. Faint curves in the background show unfiltered data whereas the filtered data are shown with thick curves.} }
    \label{fig:f3_new_round2}
\end{figure}

\textit{Discussion.---}
\bt{The turbulent cross-helicity budget, Eq.~\eqref{eq:turbcrosshel}, distinguishes the CHSI from homogeneous turbulence with uniform shear. For $U_x\propto z$ in a statistically homogeneous shearing-periodic system, the mean turbulent stress has no spatial divergence, so $T_i=\partial_jR_{ij}=0$ and the stress-production source $\langle B_iT_i\rangle_V$ is absent. Additional turbulence inhomogeneity, for example stratification, can reinstate the stress-production source by making $T_i\neq0$. Prior simulations of maintained sinusoidal shear in rotating convection \cite{HughesProctor2009,KapylaBrandenburg2010} showed large-scale magnetic-field growth above a shear threshold. In those studies, the spatial similarity between the imposed mean flow and the turbulence-generated mean magnetic field is suggestive of Eq.~\eqref{eq:BUrelation_3D} in the slow-growth or saturated limit $\lambda/(\beta_Tk^2)\ll1$. Thus, the deliberately minimal system studied here appears to be relevant to more complex shear flows with rotation, convection, or stratification.}

 \begin{figure}
    \centering
    \includegraphics[width=1\columnwidth]{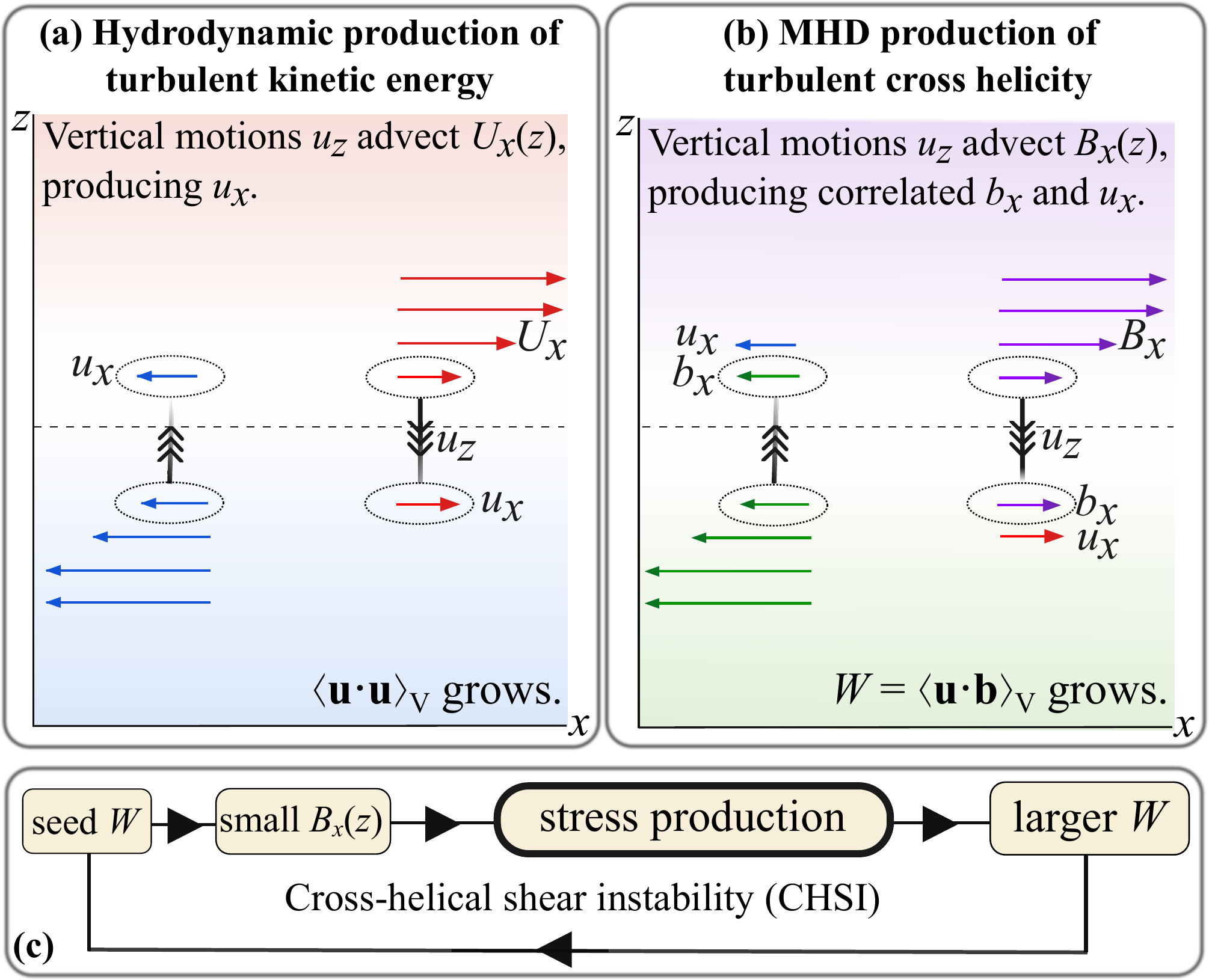}
   \caption{\bt{Physical interpretation of the cross-helical shear instability. 
   (a) In hydrodynamic shear turbulence, vertical turbulent motions $u_z$ advect the mean horizontal momentum $U_x$, producing fluctuations $u_x$. Turbulent kinetic energy grows via $P_K = \langle-R_{xz}\partial_z U_x \rangle_V$. 
   (b) In MHD, the same vertical motions advect the turbulence-generated mean magnetic field $B_x$, producing fluctuations $b_x$ aligned with $u_x$. Turbulent cross helicity grows via the stress production term $P_W = \langle-R_{xz}\partial_z B_x \rangle_V$. While $u_x$ and $b_x$ change sign along $x$ (or $y$ or both), the correlation $u_x b_x$ carries the same sign, making $W\mathrm{=}\langle \mathbf u\cdot \mathbf b\rangle_V$ non-zero and growing. 
   (c) CHSI feedback loop: a seed turbulent cross helicity $W$ generates $B_x$ through the cross-helicity EMF $\mathcal E_y \propto W $; the resulting mean field $B_x$ then enters the stress production $P_W\propto \partial_z B_x$, which acts as a source for $W$; and $W$ grows when its production exceeds losses. The CHSI does not require an initial mean magnetic field. }}
    \label{fig:sch}
\end{figure}

\bt{The cross-helical shear instability is driven by the Reynolds--Maxwell-stress gradient, $T_i\mathrm{=}\partial_jR_{ij}\mathrm{\neq}0$, as it supplies the turbulent cross-helicity source $\langle B_iT_i\rangle_V$. This source overcomes turbulent and visco-resistive losses when shear is sufficiently strong. Maintained non-uniform shear provides this stress gradient in the minimal 3D numerical setup considered here; in more general systems, other turbulence inhomogeneities can supply the stress gradient.}

\begin{acknowledgments}
\textbf{Acknowledgments:} \bt{We thank the anonymous reviewers for helpful comments that significantly improved the presentation of the work.} It is our pleasure to thank J.~R.~Beattie, A.~Bhattacharjee, E.~Blackman, E.~R.~Most, B.~Ripperda,  and A.~Philippov for fruitful discussions.  This material is based upon work funded by the National Science Foundation (NSF) under award 2409206 and Department of Energy (Grant No.~DE-SC0022257) through the DOE/NSF Partnership in Basic Plasma Science and Engineering. 
A.E.F. is supported by an NSF Astronomy and Astrophysics Postdoctoral Fellowship under award AST-2402142. 
This work used Anvil at Purdue University through allocation TG-PHY130027 from the Advanced Cyberinfrastructure Coordination Ecosystem: Services \& Support (ACCESS) program.
\end{acknowledgments}

\bibliography{refs}

\pagebreak

\newpage

\clearpage

\section*{END MATTER}

\begingroup
\color{black} % Set the text and heading color here

{\color{black}The main text evolves turbulent cross helicity $W$ dynamically through the cross helicity budget.} Here, we present the complementary quasi-stationary response closure, in which the velocity--magnetic correlator is eliminated algebraically in favor of the mean magnetic field $\mathbf B$. This gives a one-field representation of the same two-field CHSI feedback and connects the finite-shear calculations to the weak-shear Rogachevskii--Kleeorin (RK) tensor structure.

\subsection*{A.~Quasi-stationary MHD correlator response}

The fluctuations obey
\begin{subequations}
\begin{align}
D_t u_i
&=
-u_l\partial_lU_i
+B_l\partial_lb_i+b_l\partial_lB_i
+N^u_i+\nu\nabla^2u_i \label{eq:dtuiendmatter},
\\
D_t b_j
&=
-u_l\partial_lB_j
+B_l\partial_lu_j+b_l\partial_lU_j
+N^b_j+\eta\nabla^2b_j \label{eq:dtbjendmatter},
\end{align}
\end{subequations}
where $D_t \equiv \partial_t + U_l \partial_l$, and $N^u_i$ and $N^b_j$ collect nonlinear terms (including pressure). Defining
\begin{equation}
F_{ij}\equiv\langle u_i b_j\rangle ,
\end{equation}
the correlator evolves as
\begin{equation}
D_t F_{ij}=P_{ij}+T_{ij}+\epsilon_{ij},
\label{eq:MHDcorrelation_EM}
\end{equation}
where
\begin{align}
P_{ij}
&=
F_{il}\partial_lU_j
-F_{lj}\partial_lU_i
-\langle u_i u_l\rangle\partial_lB_j
+\langle b_j b_l\rangle\partial_lB_i
\nonumber\\
&\quad
+B_l\left(
\langle u_i\partial_lu_j\rangle
+\langle b_j\partial_lb_i\rangle
\right).
\label{eq:Pij_EM}
\end{align}
Here, $\epsilon_{ij}=-c_dF_{ij}/\tau$ represents visco-resistive
decorrelation, and $T_{ij}$ represents nonlinear relaxation.

In the quasi-stationary response approximation, the left-hand side of
Eq.~\eqref{eq:MHDcorrelation_EM} is set to zero. The nonlinear relaxation is
modeled in trace-preserving form
\begin{equation}
T_{ij}
=
-\frac{c_1}{\tau}
\left[
F_{ij}
-\frac{1}{3}\delta_{ij}F_{ll}
\right].
\label{eq:Tij_EM}
\end{equation}
This trace-preserving relaxation damps only the traceless part of $F_{ij}$; the trace $F_{ll}=\langle\mathbf u\cdot\mathbf b\rangle$ is preserved by the ideal nonlinear interaction, as required by the total cross helicity conservation law.

The small-scale balance $P_{ij}+T_{ij}+\epsilon_{ij}=0$ is recast as
\begin{equation}
\widehat J_{ijmn}(\mathbf U)F_{mn}
=
\widehat M_{ijk}B_k ,
\label{eq:Jbalance_EM}
\end{equation}
where
\begin{equation}
\widehat J_{ijmn}
=
C\delta_{im}\delta_{jn}
-a\delta_{ij}\delta_{mn}
+\tau\left(
\delta_{jn}\partial_mU_i-\delta_{im}\partial_nU_j
\right),
\label{eq:Joperator_EM}
\end{equation}
with \(C=c_1+c_d\) and \(a=c_1/3\), where \(c_1\) and \(c_d\) are order-unity constants. Furthermore,
\begin{equation}
\widehat M_{ijk}
=
\tau\left[
\langle u_i\partial_ku_j\rangle
-\langle u_i u_l\rangle\delta_{jk}\partial_l
+\langle b_j\partial_kb_i\rangle
+\langle b_j b_l\rangle\delta_{ik}\partial_l
\right].
\label{eq:Moperator_EM}
\end{equation}
Equation~\eqref{eq:Jbalance_EM} may be formally solved by moving the $\mathbf{U}$-dependent terms in $\widehat{J}$ to the right-hand side of Eq.~\eqref{eq:Jbalance_EM} and using $\mathcal E_i=\epsilon_{ijk} F_{jk}$, which yields a general expression
\begin{equation}\label{eq:fullemfendmatter}
\mathcal E_i = \alpha_{ij} B_j + \beta_{ijk}\partial_k B_j + \Upsilon_{ijk}\partial_k U_j,
\end{equation}
where 
\begin{subequations}
\begin{align}
\alpha_{ij}&=\widetilde{\tau} \epsilon_{ink} [\langle u_n \partial_j u_k \rangle - \langle b_n \partial_j b_k \rangle], \label{eq:endeqna}\\
\beta_{ijk}&=\widetilde{\tau} \epsilon_{ijn} [\langle u_n u_k \rangle + \langle b_n b_k \rangle], \label{eq:endeqnb}\\
\Upsilon_{ijk}&=\widetilde{\tau} \epsilon_{inj} [\langle u_n b_k \rangle + \langle u_k b_n \rangle],\label{eq:endeqnc}
\end{align}
\end{subequations}
and $\widetilde{\tau} = \tau/C$.
The last terms in Eqs.~\eqref{eq:endeqna}--\eqref{eq:endeqnc} arise from the evolution of $u_i$ in Eq.~\eqref{eq:dtuiendmatter} and the second to last terms arise from the evolution of $b_j$ in Eq.~\eqref{eq:dtbjendmatter}. When $b^2\ll u^2$, only the last terms of Eqs.~\eqref{eq:endeqna}--\eqref{eq:endeqnb} may be neglected, but not of Eq.~\eqref{eq:endeqnc}.

We note that Eq.~\eqref{eq:endeqnc} still depends on velocity--magnetic-field correlator, making the system unclosed. Hence, an alternative approach is to eliminate all velocity--magnetic-field correlators by expressing them in terms of the mean field only. This elimination procedure---the RK framework---is what we now pursue.

Hence, Eq.~\eqref{eq:Jbalance_EM} is written as 
\begin{equation}
F_{ij}
=
[\widehat J^{-1}(\mathbf U)]_{ijmn}
\widehat M_{mnk}B_k ,
\label{eq:Fformal_EM}
\end{equation}
and the EMF is
\begin{equation}
\mathcal E_i
=
\epsilon_{ijk}
[\widehat J^{-1}(\mathbf U)]_{jkmn}
\widehat M_{mnl}B_l .
\label{eq:EMFformal_EM}
\end{equation}
In the weak-shear limit $(S\ll1)$, using perturbative inversion of $\widehat J(\mathbf U)$ in Eq.~\eqref{eq:EMFformal_EM},  we have reproduced the complete RK structure of the $\beta$-tensor. It also recovers the leading structure of the $\alpha$-tensor---additional structure depends on turbulence properties. For $S\gtrsim1$, however, the inverse must be obtained nonperturbatively. For the plane-averaged flow $\mathbf U=U_x(z)\hat{\mathbf e}_x$ used below, the relevant part of $\widehat J$ reduces from a $9\times9$ operator to a $5\times5$ block.

\subsection*{B.~Plane-averaged shear and reduction of two-field model to one-field representation}

For the plane-averaged shear flow
$\mathbf U=U_x(z)\hat{\mathbf e}_x$, define the dimensionless shear
\begin{equation}
S\equiv \tau\partial_zU_x .
\end{equation}
The EMF component \(\mathcal E_y=F_{zx}-F_{xz}\) requires
\(F_{zx}\) and \(F_{xz}\). However, since $F_{zx}$ and $F_{xz}$ are coupled to other correlators in Eq.~\eqref{eq:Jbalance_EM}, one needs to solve a matrix equation involving $F_{zx}, F_{xz}, F_{xx}, F_{yy},$ and $F_{zz}$.
All other correlators decouple from $\mathcal E_y$ when the mean flow is solely $\mathbf U=U_x(z)\hat{\mathbf x}$. The
relevant part of Eq.~\eqref{eq:Jbalance_EM} is
\begin{equation}
\begin{pmatrix}
C & 0 & 0 & 0 & -S\\
0 & C & 0 & 0 & S\\
S & -S & C-a & -a & -a\\
0 & 0 & -a & C-a & -a\\
0 & 0 & -a & -a & C-a
\end{pmatrix}
\begin{pmatrix}
F_{zx}\\
F_{xz}\\
F_{xx}\\
F_{yy}\\
F_{zz}
\end{pmatrix}
=
\begin{pmatrix}
N_{zx}\\
N_{xz}\\
N_{xx}\\
N_{yy}\\
N_{zz}
\end{pmatrix},
\label{eq:five_by_five_EM}
\end{equation}
where
\begin{equation}
N_{ij}\equiv(\widehat M\mathbf B)_{ij},
\end{equation}
contains the source terms due to the mean magnetic field and its gradients. 

For the plane-averaged mean field with $B_z=0$, the sources relevant to the CHSI branch (nonhelical) are
\begin{align}
N_{zx}
&=
-\tau\langle u_z^2\rangle\,\partial_zB_x,
&
N_{xz}
&=
+\tau\langle b_z^2\rangle\,\partial_zB_x,
\\
N_{xx}
&=
-\tau R_{xz}\partial_zB_x,
&
N_{yy}
&=
-\tau R_{yz}\partial_zB_y,
&
N_{zz}
&=0,
\end{align}
where
\begin{equation}
R_{iz}\equiv\langle u_i u_z-b_i b_z\rangle .
\end{equation}
Note that the source $N_{xx}$ is the coupling between magnetic-field gradient and the Reynolds--Maxwell stress, which dominates $N_{yy}$  as $R_{xz}>R_{yz}$, for $U_x(z)$. Thus, $N_{xx}$ is the exact same coupling that re-generates turbulent cross helicity and leads to the CHSI in the main text of this Letter; see the second term in Eq.~\eqref{eq:turbcrosshel}.

Using exact inversion of Eq.~\eqref{eq:five_by_five_EM} and $\mathcal E_y
=
F_{zx}-F_{xz}$
\begin{equation}
\mathcal E_y
=
-\tau
\frac{
C c_d\langle u_z^2\mathrm{+}b_z^2\rangle\,\partial_zB_x
+ 2aS\left(
R_{xz}\partial_zB_x\mathrm{+}R_{yz}\partial_zB_y
\right)
}{
C^2c_d+2aS^2
}.
\label{eq:Ey_exact_EM}
\end{equation}

The stress $R_{xz}$ dominates over $R_{yz}$ because the maintained mean
momentum is directed along $x$ and varies along $z$,
$\mathbf U=U_x(z)\hat{\mathbf e}_x$. Hence Eq.~\eqref{eq:Ey_exact_EM}
reduces to
\begin{equation}
\mathcal E_y=-\eta_{\rm eff}(S)\partial_zB_x ,
\label{eq:Ey_etaeff_EM}
\end{equation}
where
\begin{equation}
\eta_{\rm eff}(S)
=
\tau
\frac{
C c_d\langle u_z^2+b_z^2\rangle
- 2a\times(-S R_{xz})
}{
C^2c_d+2aS^2
}.
\label{eq:etaeff_exact_EM}
\end{equation}
This $\eta_{\rm eff}$ is {\em not} an independent diagonal-$\beta$
coefficient. It is the reduced transport representation of the same
stress--cross-helicity feedback described dynamically in the main text. For
$S=0$, Eq.~\eqref{eq:etaeff_exact_EM} reduces to the ordinary positive
turbulent diffusivity, up to the order-unity coefficient $C$. For down-gradient
momentum transport, $SR_{xz}<0$. Therefore, the term $-SR_{xz}$ can reduce
the effective diffusivity, making it negative, in regions of finite shear.

It is strongly emphasized that the expression for $\mathcal{E}_y$ in Eq.~\eqref{eq:Ey_etaeff_EM} is exactly the same as $\mathcal{E}_y = \Upsilon_{yxz}\partial_z U_x + \beta_{yxz}\partial_z B_x$. To confirm this identity, we use Eq.~\eqref{eq:endeqnc} to derive $\Upsilon_{yxz} = 2\widetilde{\tau} \langle u_z b_z\rangle$ and $\beta_{yxz}=-\widetilde{\tau}\langle u_z^2 + b_z^2\rangle$. The velocity--magnetic-field correlator $F_{zz}=\langle u_z b_z\rangle$ is then derived in terms of the mean field by using the last row of Eq.~\eqref{eq:five_by_five_EM}, when the left-hand side matrix is inverted. This procedure gives $
\Upsilon_{yxz} = -2 a \tau S \partial_z B_x  (\beta_{yxz}+\tau R_{xz}S^{-1}) /(C^2 c_d + 2 a S^2).$

Estimating the down-gradient momentum transport measured in the kinematic phase $
R_{xz}
\simeq
-\mu_T\partial_zU_x
=
-\mu_T S/\tau$,
Eq.~\eqref{eq:etaeff_exact_EM} becomes
\begin{equation}
\eta_{\rm eff}(S)
=
\frac{
\tau C c_d\langle u_z^2+b_z^2\rangle
-2a\mu_T S^2
}{
C^2c_d+2aS^2
}.
\label{eq:etaeff_down_gradient_EM}
\end{equation}
The weak-shear limit $S\ll1$ is diffusive as
\begin{equation}
\eta_{\rm eff}(S)
=
\frac{\tau}{C}\langle u_z^2+b_z^2\rangle
\left[1{\color{black}{-O(S^2)}}\right]>0.
\end{equation}
Hence, while the turbulent stress attempts to make $\eta_{\rm eff}$ less positive (less diffusive) via ${\color{black}{-O(S^2)}}$ term even in the weak-shear regime, this effect  is subdominant. This is consistent with
the weak-shear ordering of the RK theory, in which the stress-production effect is not sufficiently strong to destabilize the plane-averaged
mean-field considered here.
In the finite-shear regime, however, the stress correction can overcome the
positive turbulent and molecular diffusion, producing an instability of $B_x$.

With finite molecular resistivity, the reduced one-field problem is the
global eigenvalue problem
\begin{equation}
\partial_tB_x
=
\partial_z\left[
\left\{\eta+\eta_{\rm eff}(z)\right\}\partial_z B_x
\right].
\end{equation}
Thus, the condition for the growth of $B_x$ --- and the growth of $\langle \mathbf u \cdot \mathbf b \rangle \propto \partial_z B_x$ --- is determined by the global operator, not by a
pointwise criterion. For qualitative insight, one may consider the ideal local estimate
$\eta_{\rm eff}<0$ 
\begin{equation}
S
\gtrsim
\sqrt{ \frac{3c_d \beta_z}{\mu_T}},
\label{eq:etaeff_threshold_EM}
\end{equation}
where $\beta_z=\tau\langle u_z^2+b_z^2\rangle$ denotes the vertical anisotropic contribution to the turbulent diffusivity. The intermediate factor $C/(2c_1)\approx 1$, as $c_1$ and $c_d$ are of order unity.

For homogeneous linear shear, this reduced one-field model should not be interpreted as the instability of $\langle \mathbf u\cdot \mathbf b\rangle_V$. If \(U_x \propto z\) and turbulence is statistically homogeneous, then $\langle \mathbf u\cdot \mathbf b\rangle_V = 0$. This is because \(R_{ij}\) is spatially uniform, \(T_i=\partial_jR_{ij}=0\), and the stress-production source \(\langle B_iT_i\rangle_V\) is absent. The growth of $\langle \mathbf u\cdot \mathbf b\rangle_V$ requires a stress gradient: either $S$ or $\mu_T$ should vary in space.

Thus, the theoretical expectation of the above calculation is that the stress-production effect is subdominant when
$S\ll1$, but can overcome losses when $S \gtrsim 1$. The criterion~\eqref{eq:etaeff_threshold_EM} agrees with the criteria~\eqref{eq:Sconditon} and~\eqref{eq:QLestimate}. All three mean-field derivations---cross helicity budget, quasilinear, and nonlinear closures---produce a consistent prediction.

\endgroup

\pagebreak

\clearpage

\newpage

\clearpage
\onecolumngrid

\setcounter{page}{1}
\renewcommand{\thepage}{SI-\arabic{page}}

\setcounter{equation}{0}
\renewcommand{\theequation}{S\arabic{equation}}

\setcounter{figure}{0}
\renewcommand{\thefigure}{S\arabic{figure}}

\section*{Supplementary Information (SI) for ``Exponential Growth of Turbulent Cross Helicity in Shear-Flow-Driven Turbulence''}

\subsection*{S1.~Derivation of the cross helicity budget}
Considering an arbitrary ensemble average denoted by $\langle \cdot\rangle$, \[
\mathbf u_{\rm tot}=\mathbf U+\mathbf u,\qquad 
\mathbf b_{\rm tot}=\mathbf B+\mathbf b,
\]
where $\mathbf U=\langle\mathbf u_{\rm tot} \rangle$ and $\mathbf B=\langle\mathbf b_{\rm tot} \rangle$ represent large-scale (mean) flow and magnetic field. Then, the fluctuations $\mathbf u$ and $\mathbf{b}$ in incompressible magnetohydrodynamics obey
\begin{subequations}
\begin{align}
D_t u_i
&=
-u_l\partial_lU_i
+B_l\partial_lb_i+b_l\partial_lB_i
-u_l \partial_l u_i + b_l \partial_l b_i -\partial_i p_\ast +\nu\nabla^2u_i \label{eq:dtuiendmatterSI},
\\
D_t b_i
&=
-u_l\partial_lB_i
+B_l\partial_lu_i+b_l\partial_lU_i
-u_l \partial_l b_i + b_l \partial_l u_i +\eta\nabla^2b_i \label{eq:dtbjendmatterSI},
\end{align}
\end{subequations}
where $D_t \equiv \partial_t + U_l \partial_l$. No direct external forcing is applied to fluctuations in this work.

We multiply Eq.~\eqref{eq:dtuiendmatterSI} with $b_i$ and Eq.~\eqref{eq:dtbjendmatterSI} with $u_i$ and add them together, after which the resulting equation is volume-averaged $\langle\cdot\rangle_V$ to obtain
\begin{align}\label{eq:SI1}
\partial_t  \langle u_i b_i \rangle_V +   \underbrace{\langle \partial_l (U_l u_i b_i) \rangle_V}_{{\rm FD}}
&=
 \Bigg\langle b_i\left(-u_l \partial_lU_i -\underbrace{u_l \partial_l u_i}_{\rm ADD} +\underbrace{B_l\partial_lb_i}_{{\rm FD}}+b_l\partial_lB_i + \underbrace{b_l \partial_l b_i}_{{\rm FD}} +\nu\nabla^2u_i-\underbrace{\partial_i p_\ast}_{{\rm FD}} \right)\Bigg\rangle_V \nonumber\\
&+ \Bigg\langle u_i\left(-u_l\partial_lB_i -\underbrace{u_l \partial_l b_i}_{\rm ADD} +\underbrace{B_l\partial_lu_i}_{{\rm FD}}+b_l\partial_lU_i + \underbrace{b_l \partial_l u_i}_{{\rm FD}} +\eta\nabla^2b_i\right)\Bigg\rangle_V.
\end{align}
Using $\partial_i u_i = \partial_i b_i = 0$, the terms labeled with ${\rm ``FD"}$ in Eq.~\eqref{eq:SI1} can be directly written as flux divergences, which upon volume-averaging become zero, when suitable boundary conditions are used. The terms labeled with ${\rm ``ADD}"$ in  Eq.~\eqref{eq:SI1} are added to write $-\langle u_l \partial_l (u_i b_i) \rangle_V$, which represents advection of  $u_i b_i$. Since this additive term also can be written as a flux divergence $-\langle  \partial_l ( u_l u_i b_i) \rangle_V$, it  also becomes zero upon volume-averaging. Hence, the resulting equation is 
\begin{align}
\partial_t  \langle u_i b_i \rangle_V 
&= \langle b_i\left(-u_l \partial_lU_i +b_l\partial_lB_i +\nu\nabla^2u_i \right)\rangle_V \nonumber\\
&+ \langle u_i\left(-u_l\partial_lB_i  +b_l\partial_lU_i   +\eta\nabla^2b_i\right)\rangle_V,
\end{align}
which can be compactly written as
\begin{equation}\label{eq:turbcrosshel3}
\partial_t  \langle \mathbf u \cdot \mathbf b\rangle_V \mathrm{=} 
\hspace{0.2cm}\underbrace{\langle \mathbf{B}\cdot\mathbf{T}\rangle_V}_{\textrm{\hspace{-1.0cm}}\substack{\text{\BT{Stress-production}}\\ \text{\BT{}}}\hspace{-1cm}} 
-(\nu+\eta)\langle\boldsymbol{\omega}\cdot\mathbf{j}\rangle_V 
\underbrace{- \langle \mathbf{U}\cdot\nabla\times\boldsymbol{\mathcal{E}}\rangle_V}_{\textrm{\hspace{-1.0cm}}\substack{\text{\BT{EMF transfer}}\\ \text{\BT{}}}\hspace{-1cm}},
\end{equation}
where $T_i \mathrm{=} \partial_j R_{ij}$ and $R_{ij}\mathrm{=}\langle u_i u_j-b_i b_j\rangle$ represents the Reynolds--Maxwell stress. Using suitable boundary conditions,  $\langle \mathbf{B}\cdot\mathbf{T}\rangle_V=-\langle R_{ij}\partial_j B_i\rangle_V$ and $\langle \mathbf{U}\cdot\nabla\times\boldsymbol{\mathcal{E}}\rangle_V=\langle \boldsymbol{\mathcal{E}}\cdot(\nabla\times\mathbf{U})\rangle_V$.

Equation~\eqref{eq:turbcrosshel3} is valid for incompressible MHD with arbitrary large-scale flow \(\mathbf U\) and magnetic field \(\mathbf B\), subject to the boundary conditions used above. Since the fluctuations are not directly forced externally, external forcing term is absent in Eq.~\eqref{eq:turbcrosshel3}. In the absence of a large-scale flow ($\mathbf U=0$, which naturally implies $\mathbf T=0$), turbulent cross helicity is depleted by visco-resistive effects. In the presence of a large-scale flow, however, the budget \eqref{eq:turbcrosshel3} alone does not determine the sign of the evolution: whether \(\langle\mathbf u\cdot\mathbf b\rangle_V\) decays or grows depends on the turbulence properties measured by \(\boldsymbol{\mathcal E}\) and \(\mathbf T\), and on their couplings to \(\mathbf U\) and \(\mathbf B\). Thus, even when a large-scale flow \(\mathbf U\)  or magnetic field \(\mathbf B\)  is prescribed and sustained, whether turbulent cross helicity grows or decays is not known {\em a priori}. This is where the present work brings contribution. We show in this work that when the Reynolds--Maxwell-stress gradient $T_i = \partial_j R_{ij}$ is non-zero and the shear is strong, the couplings of $\boldsymbol{\mathcal E}$ and $\mathbf T$ with the mean flow $\bf U$ and the mean magnetic field $\bf B$ enable the stress-production term to overcome the effective losses. The result is the growth of $\langle \mathbf u \cdot \mathbf b\rangle_V$.

\subsection*{S2.~ Comparison of the cross helicity budget with the hydrodynamic turbulent kinetic energy budget}
It can be instructive to draw an analogy of Eq.~\eqref{eq:turbcrosshel3} with the well-known production of turbulent kinetic energy in hydrodynamics: replacing ${\mathbf{b}}$ with ${\mathbf{u}}$ on the left-hand side of Eq.~\eqref{eq:turbcrosshel3}, we derive an evolution equation of turbulent kinetic energy in hydrodynamic shear flow
\begin{equation}
 \frac{\partial_t\langle{\mathbf{u}}\cdot{\mathbf{u}} \rangle_V}{2} =  \langle \mathbf{U}\cdot\mathbf{T}\rangle_V - \nu \langle {\boldsymbol{\omega}}\cdot{\boldsymbol{\omega}}\rangle_V.
\end{equation}

The hydrodynamic turbulence production term $\langle \mathbf{U}\cdot\mathbf{T}\rangle_V=-\langle R_{ij}\partial_j U_i\rangle_V$ 
transfers energy from mean to fluctuations. This is because the down-gradient transport of momentum enforces the relation $-\langle R_{ij} \partial_j U_i \rangle_V \propto \langle \partial_j U_i \partial_j U_i \rangle_V >0$.  However, the MHD term $\langle \mathbf{B}\cdot\mathbf{T}\rangle_V$ is not sign-definite and it is {\em a priori} unknown whether it supplies or removes turbulent cross helicity, even when the shear  is sustained or its profile is known. Additionally, our simulations start with zero mean magnetic field $\mathbf B = 0$, and hence $\langle \mathbf{B}\cdot\mathbf{T}\rangle_V$ is zero in the beginning. Thus, any role of $\langle \mathbf{B}\cdot\mathbf{T}\rangle_V$ later on is entirely due to turbulence. This work shows that shear-flow turbulence has a peculiar tendency:  while $\langle \mathbf{B}\cdot\mathbf{T}\rangle_V$ has no definite sign, the CHSI operates such that $\langle \mathbf{B}\cdot\mathbf{T}\rangle_V$, on average, carries the  same sign as that of turbulent cross helicity $W=\langle\mathbf u\cdot \mathbf b \rangle_V$, thus amplifying $W$, when production exceeds losses. The losses are due to the EMF transfer term and visco-resistive effects.  The role of a finite shear is to enhance the production of $W$ relative to turbulent and diffusive losses. When production exceeds losses, $W$ grows. However, the nature of growth, e.g., linear or exponential, is not determined by the budget equation \eqref{eq:turbcrosshel3} alone. The exponential growth of $W$ is a consequence of the CHSI feedback, shown in Fig.~\ref{fig:sch}(c).

\subsection*{S3.~Derivation of the CHSI dispersion relation}
Here, a step-by-step derivation of the CHSI dispersion relation is presented. We write each term of Eq.~\eqref{eq:turbcrosshel} in terms of turbulent cross helicity $W$. 

The first term on the right-hand side of Eq.~\eqref{eq:turbcrosshel} is
\begin{equation}
\langle \mathbf{B}\cdot \mathbf{T}\rangle_V = \sum_{\mathbf{k}} \widehat{\mathbf{B}}^\ast(\mathbf{k}) \cdot\widehat{\mathbf{T}}(\mathbf{k}) \mathrm{=}\frac{2\tau W}{3} \sum_{\mathbf{k}} \frac{k^2 \widehat{\mathbf{U}}^\ast(\mathbf{k})}{\lambda+\beta_\mathrm{T} k^2} \cdot\widehat{\mathbf{T}}(\mathbf{k}),
\end{equation}
where Eq.~\eqref{eq:BUrelation_3D} is used. The asterisk denotes complex conjugation. 

The second term on the right-hand side of Eq.~\eqref{eq:turbcrosshel} is written as 
\begin{equation}
(\nu+\eta) \langle {\boldsymbol{\omega}}\cdot{\mathbf{j}}\rangle_V  =  \frac{\langle {\mathbf{u}}\cdot{\mathbf{b}}\rangle_V}{\tau_{\rm diss}} = \frac{W}{\tau_{\rm diss}}. \label{eq:S7}
\end{equation}

When cross helicity provides dominant source in the turbulent EMF, the last term of Eq.~\eqref{eq:turbcrosshel}  becomes
\begin{align}
\langle \mathbf{U}\cdot\nabla\times\boldsymbol{\mathcal E}\rangle_V 
&=\langle (\nabla\times\mathbf{U})\cdot\boldsymbol{\mathcal E}\rangle_V \nonumber\\
&=\Big\langle (\nabla\times \mathbf U)\cdot\left[\frac{2\tau W}{3} \nabla\times \mathbf U - \beta \nabla\times\mathbf{B}\right]\Big\rangle_V \nonumber\\
&=\frac{2\tau W}{3} \sum_\mathbf{k} k^2 |\widehat{\mathbf{U}}(\mathbf{k})|^2 - \frac{2\tau W}{3} \sum_\mathbf{k} \frac{\beta k^4 |\widehat{\mathbf{U}}(\mathbf{k})|^2 }{\lambda +\beta_\mathrm{T} k^2}  \nonumber\\
&=\frac{2\tau W}{3}  \sum_{\mathbf{k}}\,\,\, \left[\frac{k^2 |\widehat{\mathbf{U}}(\mathbf{k})|^2 (\lambda +\eta k^2)}{\lambda +\beta_\mathrm{T} k^2}\right], \label{eq:S8}
\end{align}
where $\mathbf{B}$ is expressed in terms of $\mathbf{U}$ using Eq.~\eqref{eq:BUrelation_3D}. When $\lambda >0$ (the cross-helical shear instability), $\mathcal P_{\rm EMF} = -\langle \mathbf{U}\cdot\nabla\times\boldsymbol{\mathcal E}\rangle_V $ is negative relative to $W$. Figures 3 and 4 further confirm that the term $-\langle \mathbf{U}\cdot\nabla\times\boldsymbol{\mathcal E}\rangle_V $ transfers turbulent cross helicity from fluctuations to the mean fields. The physical meaning of the negative $\mathcal P_{\rm EMF}$ (relative to $W$) is simple: When turbulent cross helicity generates a mean magnetic field $\bf B$ via the cross-helicity EMF, that $\bf B$ tends to be aligned with the mean flow $\bf U$, say $\mathbf B || \mathbf U$ in a particular case. This non-zero cross helicity in the mean field  is a result of transfer of cross helicity from fluctuations to the mean field via the cross-helicity EMF. Hence, the negative $\mathcal P_{\rm EMF}$ denotes a loss for fluctuations (and a gain for the mean fields). This loss for fluctuations is exactly what is captured by the blue curves in Figs.~3 and 4.

Adding Eqs~.\eqref{eq:S7} and \eqref{eq:S8},
\begin{equation}\label{eq:cdampSI2}
\langle \mathbf{U}\cdot\nabla\times\boldsymbol{\mathcal E}\rangle_V + (\nu+\eta) \langle {\boldsymbol{\omega}}\cdot{\mathbf{j}}\rangle_V = W c_{\rm damp}(\lambda).\end{equation} 
The expression for $c_{\rm damp}(\lambda)$ is given in Eq.~\eqref{eq:cdamp} in the main text.

Thus, when one writes a spectral representation of Eq.~\eqref{eq:turbcrosshel},  $W$ cancels out from each term, yielding the CHSI dispersion relation
\begin{equation}\label{eq:dispersionSI}
\begin{aligned}
\lambda+ c_{\rm damp}(\lambda)= \frac{2\tau}{3}  \sum_{\mathbf{k}}\,\,\, \left[\frac{k^2 \widehat{\mathbf{U}}^\ast(\mathbf{k})\cdot \widehat{\mathbf{T}}(\mathbf{k})}{\lambda +\beta_\mathrm{T} k^2}\right],
\end{aligned}
\end{equation} 
where the term on the right-hand side is the CHSI source. This source arises from the stress-production in Eq.~\eqref{eq:turbcrosshel}. The term  \(c_{\rm damp}(\lambda)\) represents the rate of effective losses of turbulent cross helicity due to visco-resistive effects and the EMF transfer terms. 

\subsection*{S4.~CHSI growth rate vs.~simulation parameters}
For most runs, we simulate a domain of size $L_x\times L_y\times L_z$, with $L_x/a=L_y/a=10\pi$ and $L_z/a=20\pi$. We use up to $1024\times 1024\times 2048$ spatial resolution with a pseudospectral solver GHOST \cite{alexakis2005,mininni2005}. 

We have also simulated cases of different domain sizes, boundary conditions (periodic and non-periodic), initial conditions, and system parameters, and have reproduced the CHSI. We varied $L_y$ in distinct simulations from $10\pi$ to $20\pi, 40\pi, 80\pi, 160\pi,$ and $320\pi$, with correspondingly large number of grid points. For the case of $L_y=320\pi$, we used 4096 grid points in $y$. The growth rate of the CHSI is shown for critical parameters in Fig.~\ref{fig:f5_app} 

\subsection*{S5.~Code benchmark and parallelization}
We parallelize the pseudospectral solver GHOST \cite{alexakis2005,mininni2005} using MPI in $z$ and openMP in $y$ to leverage the non-uniform memory-access infrastructure for an efficient solution for large-resolution simulations. We have benchmarked GHOST against another pseudospectral solver, Dedalus \cite{burns2020,lecoanet2016}.

 \begin{figure*}
    \centering
    \includegraphics[width=0.9\columnwidth]{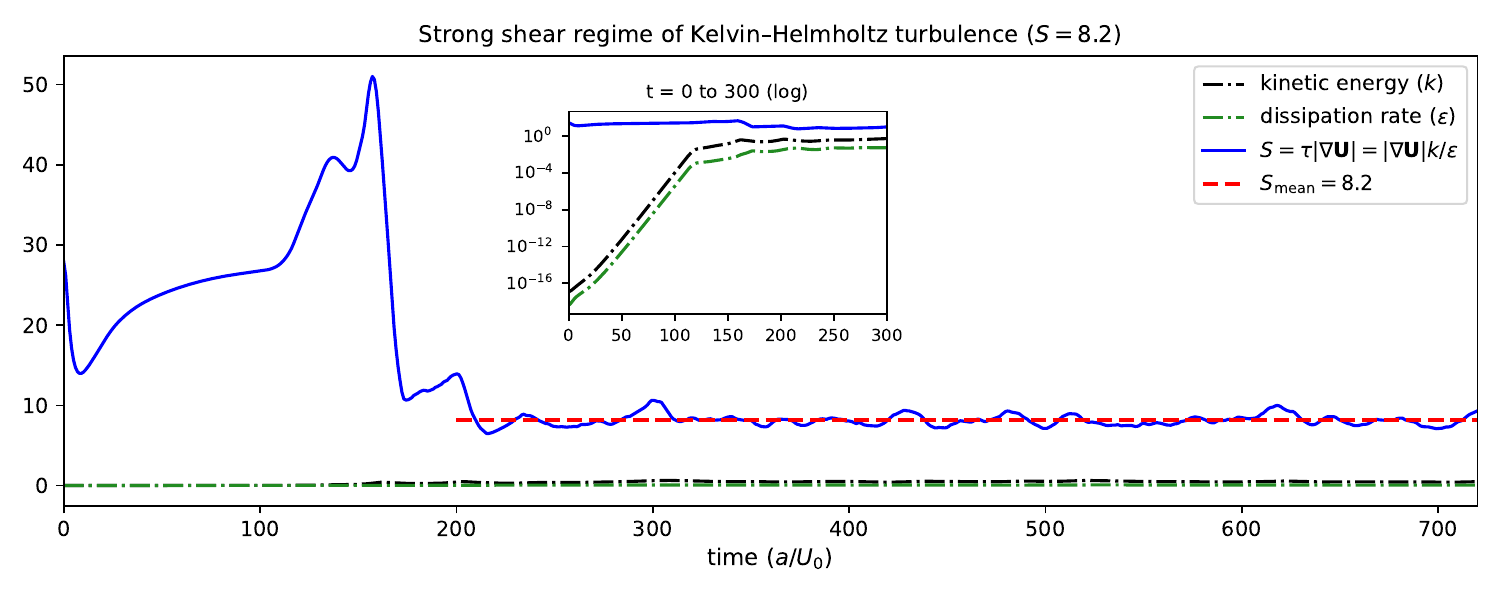}
    \caption{\bt{The KHI of the maintained shear flow drives turbulence that operates in the finite-shear regime, $S\sim \mathcal{O}(10)$. During the initial transient phase (up to $t \approx 200 a/ U_0$), flow fluctuations grow exponentially via KHI and become nonlinearly re-organized into saturated 3D structure. At this stage, the rate of energy injection ($k/\tau$) by the Kelvin--Helmholtz instability approximately balances the rate of viscous dissipation ($\epsilon$), thus setting up a quasi-stationary hydrodynamic turbulence. Then, the kinematic CHSI ensues.}}
    \label{fig:shear_rate}
\end{figure*}

 \begin{figure*}
    \centering
    \includegraphics[width=0.6\columnwidth]{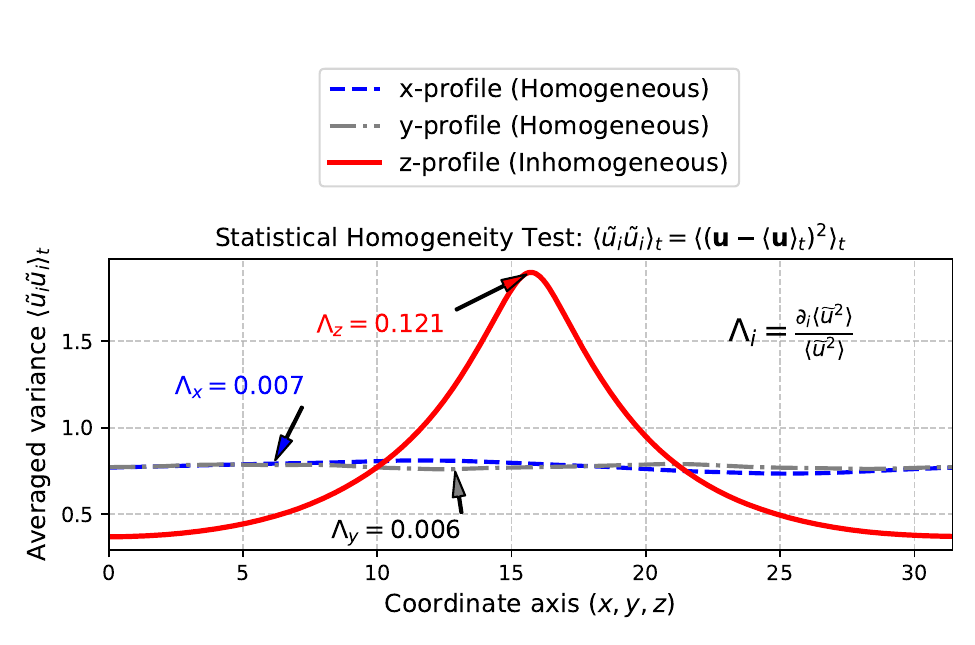}
    \caption{\bt{The directions $x$ and $y$ are statistically homogeneous, whereas the shear-wise direction $z$ is statistically inhomogeneous.  In this figure, $\langle \cdot\rangle_t$ denotes temporal (ensemble) average over the saturated stage of flow turbulence; thus, $\langle u_i\rangle_t$ denotes the time-average of the total flow $u_i$ in the system. The residual $\tilde{u}_i  = u_i- \langle u_i\rangle_t$  features a full 3D spatial profile at each time. The symbol $\langle \tilde{u}_i\tilde{u}_i\rangle_t$ denotes the time-averaged 3D spatial profile of turbulent energy. From this quantity, inverse length scale $\Lambda_i$ of turbulence intensity is determined along each direction $i\in \{x,y,z\}$. Informed by this figure, {\em mean} is defined in simulations by averaging variables along the statistically homogeneous directions $x$ and $y$. }}
    \label{fig:homogeneity_test}
\end{figure*}

 \begin{figure*}
    \centering
    \includegraphics[width=0.65\columnwidth]{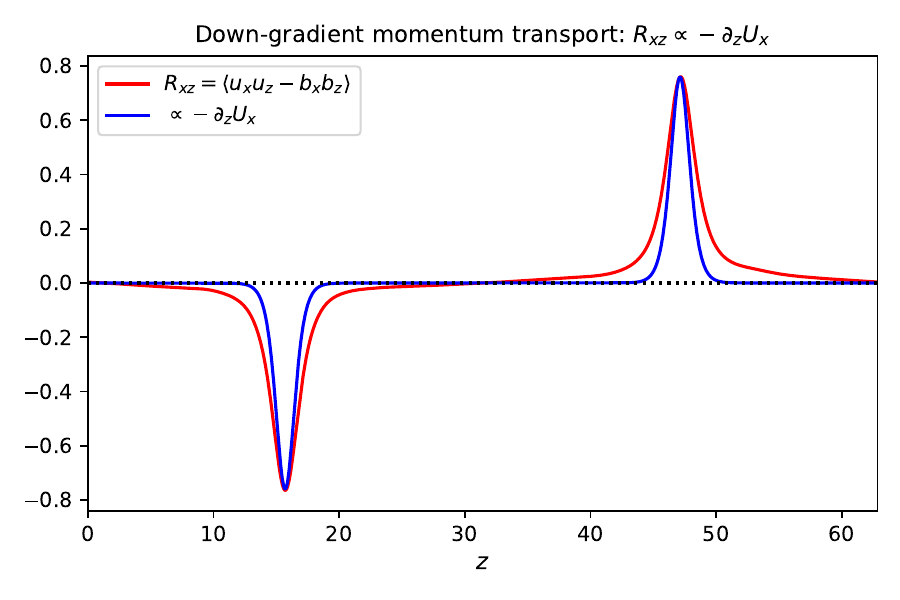}
    \caption{\bt{The spatial profiles of the Reynolds--Maxwell stress $R_{xz}$ are time-averaged over the CHSI phase ($t=200\textrm{--}1000$). The similarity between the profiles of $R_{xz}$ and $-\partial_z U_x$ indicates that turbulent momentum flux follows, to a good degree, the gradient-transport form in the simulation where the mean shear flow is maintained. This simulation with grid resolution of $512\times 512\times 1024$ is the same as the one shown in Fig.~\ref{fig:f2_new}.}}
    \label{fig:turbulent_stress_profile}
\end{figure*}

 \begin{figure*}
    \centering
    \includegraphics[width=0.5\columnwidth]{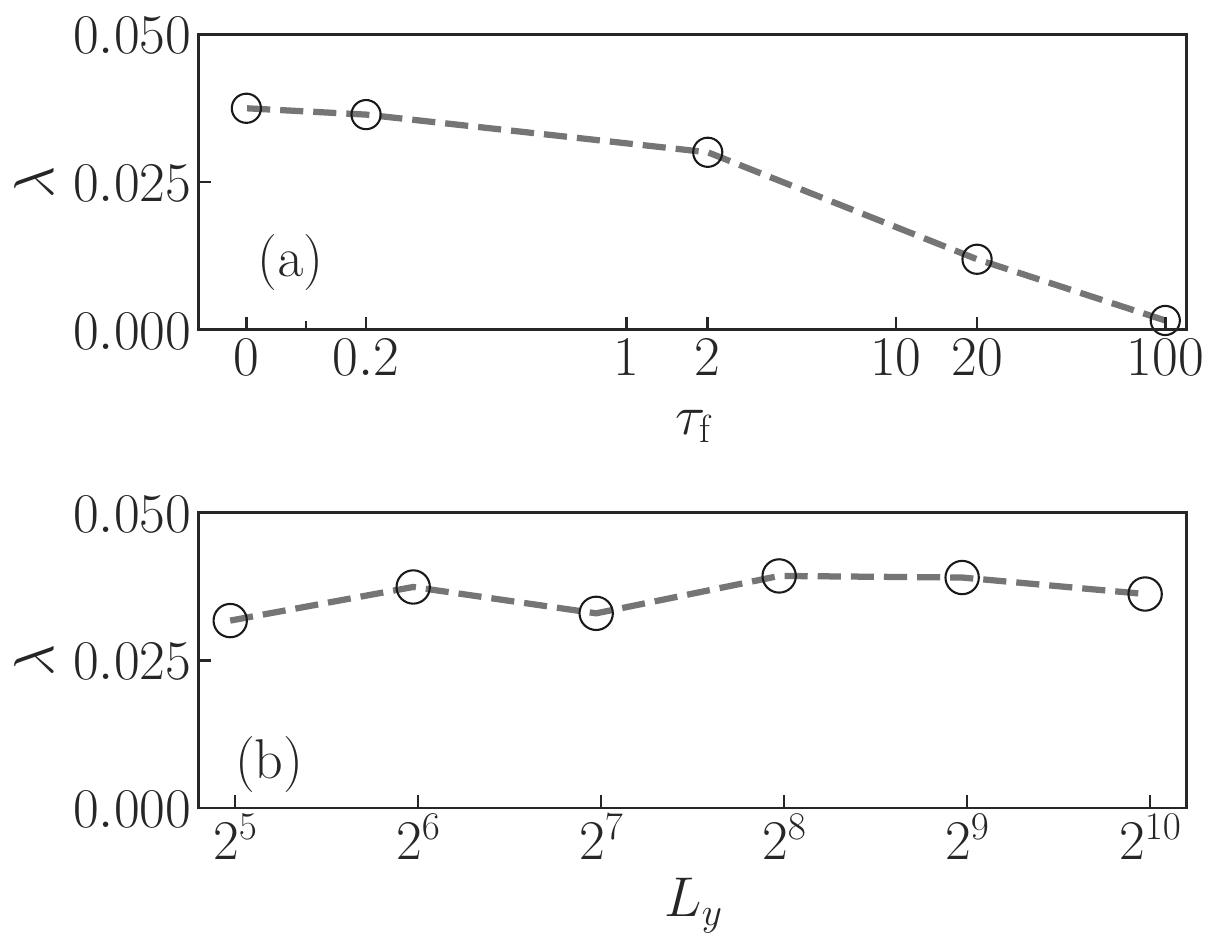}
    \caption{(a)  The growth rate of turbulent cross helicity is measured in 3D MHD simulations by varying the time-scale  $\tau_\mathrm{f}$ over which the mean shear flow is dynamically driven back to its initial unstable equilibrium. The parameter $\tau_{\rm f}=0$ represents the case where the mean shear flow is frozen in time. So long as the mean flow is maintained faster than it is depleted by the Kelvin--Helmholtz instability, the CHSI growth is observed.     (b) The CHSI growth rate remains unchanged when the domain  size $L_y/a \gg 1$ is further enlarged. }
    \label{fig:f5_app}
\end{figure*}

\clearpage

\end{document}